\documentclass[floatfix,secnumarabic,amssymb,nobibnotes,nofootinbib,aps,pra,showpacs]{revtex4}
\usepackage{epsfig}
\usepackage[colorlinks=true, citecolor=blue, linkcolor=blue, urlcolor=black]{hyperref}
\usepackage{cleveref}
\usepackage{times}
\usepackage{amssymb}
\usepackage{amsmath}
\usepackage{mathrsfs}
\usepackage{graphicx}
\usepackage{epsfig}
\usepackage{dcolumn}
\usepackage{color}
\usepackage{bm}
\usepackage{graphics,psfrag}
\usepackage{graphicx,psfrag}
\usepackage{soul}

\begin{document}
\setlength{\topmargin}{0in}
\title{Suppressing deleterious effects of spontaneous emission in creating bound states in cold atom continuum}
\author{Somnath Naskar$^{1,2}$}\thanks{These authors contributed equally to this work.} \author{Dibyendu Sardar$^1$}\thanks{These authors contributed equally to this work.} \author{Bimalendu Deb$^{1}$ and G. S. Agarwal$^3$}
\affiliation{$^{1}$School of Physical Sciences, Indian Association for the Cultivation of Science (IACS), Jadavpur, Kolkata 700032, INDIA.
\\ $^2$Department of Physics, Jogesh Chandra Chaudhuri College, Kolkata-700033, India. \\$^3${Institute for Quantum Science and Engineering}, Departments of Biological and Agricultural Engineering and Physics and Astronomy, Texas A\&M University, College Station, Texas 77843, USA}

\begin{abstract}
In a previous paper  [B. Deb and G. S. Agarwal, Phys. Rev. A  {\bf 90}, 063417 (2014)], it was theoretically shown that, magneto-optical manipulation of low energy scattering resonances and atom-molecule transitions could lead to the formation of a bound state in continuum (BIC), provided there is no spontaneous emission.  We find that  even an exceedingly small spontaneous decay from exited molecular states can spoil the BIC. In this paper, we show how to circumvent the detrimental effect of spontaneous emission by making use of vacuum-induced coherence (VIC) which results in the cancellation or suppression of spontaneous emission. VIC occurs due to the destructive  interference  between two spontaneous decay pathways. An essential condition for VIC is the non-orthogonality of the transition dipole moments associated with the decays. Furthermore, the interference between decay pathways requires that the spacing between the two decaying states must be comparable to or smaller than the square root of the product of the two 
spontaneous linewidths. We demonstrate that these conditions can be fulfilled by microwave dressing of two appropriately chosen molecular excited states, opening a promising prospect for the  experimental realization of BIC of cold atoms.
\end{abstract}
\maketitle

\section{introduction}
The idea of a bound state in continuum (BIC) was first put forward by J. von Neuman and E. Wigner in 1929 \cite{neuman1929}, in very early days of quantum mechanics. Such a state is particularly unusual and counter-intuitive as it is a localised square-integrable state despite the energy eigenvalue of the state being above the continuum threshold for the usual negative-energy bound states. To create such a state, Neuman and Wigner considered the amplitude modulation of a free-particle wave function, resulting in  a local potential which could support a BIC.  Since then, this idea of local potential was put under extensive theoretical investigations in a number of works \cite{jain1975,gazdy1977,molina2012}. Two-particle BIC has been shown to have a deep connection with  two-body resonance scattering theory \cite{fonda1960,friedrich1985a,friedrich1985b} where BIC is recognised to be related to a resonance of zero width. Owing to its fundamental significance in quantum mechanics, BIC has been re-examined by several 
authors  \cite{robnik1986,nockel1992,duclos2001,ordonez2006,prodanovic2006,weimann2013,rivera2016}.  The fact that the underlying concept in BIC pertains to wave phenomenon has motivated several theoretical and experimental works towards its possible realization and applications in  various types of material systems such as superlattices \cite{herrick1977,stillinger1977}, photonic crystal structures \cite{marinica2008,hsu2013b}, acoustic waves \cite{porter2005,linton2007}, two-particle Hubbard model \cite{zhang2012}, optical systems \cite{capasso1992} and so on. In recent times, experimental realisation of BIC in extended photonic structures is reported by  several workers \cite{plotnik2011,hsu2013a,kodigala2017}. BIC has found an important application in creating an exotic laser system called  BIC laser \cite{kodigala2017}. There is an excellent review in literature written by Hsu et. al. \cite{hsu2016}, where the study of BIC is categorically summarized on the basis of different theoretical approaches and 
corresponding experimental implementations. 

In the context of atomic and molecular systems, BIC is yet to be realized. In the past, the possibility of creating a BIC in atomic and molecular systems was discussed by several workers  \cite{friedrich1985a,stillinger1974,stillinger1975}. With the recent advent of high-precision  spectroscopy and coherent control of ultracold atoms and molecules, the realization of a BIC for ultracold atoms appears to be promising.   Recently, the creation of a BIC by magneto-optically controlling ultracold collisions of atoms has been  proposed by two of us \cite{deb2014}. This proposal has made use of two excited molecular states without considering any spontaneous emission from these states. A generic feature of atomic or molecular system is the spontaneous decay of excited states into lower-energy states via vacuum-field induced electric dipole transitions. 

In this paper, we study the effects of spontaneous emission on the atomic BIC as proposed in the Ref.  \cite{deb2014}. Our model system consists of three molecular bound states which interact with  the continuum states of ground-state atom-atom collisions in the presence of external magnetic and optical fields. To include the spontaneous emission in the model, we consider that the continuum of vacuum electromagnetic modes  interact with the excited molecular states. Our results show that spontaneous emission is a tremendous hindrance to the forming of the proposed BIC. In this work, we show how to overcome this challenge by almost completely nullifying the detrimental effect of spontaneous emission by making use of vacuum induced coherence (VIC) \cite{agarwal_book}.  We show that, VIC {\mbox{\cite{agarwal_book,Fleischhauer1992,zhu1995,zhu1996,das2012,ficek_book}}} which is basically an interference phenomenon between two spontaneous emission pathways,  can  play a decisive role in the formation of BIC in a realistic model that we present in this work. 

One of the key conditions for VIC to take place is the non-orthogonality of the dipole moments involved in two spontaneous emission pathways. Another essential condition is that the two decaying excited states should be energetically close enough compared to the geometric mean of the two spontaneous emission linewidths. Coherently controlled ro-vibrational level structure of an excited  molecule driven by a pair of photoassociation (PA) lasers has been shown to  facilitate for the fulfillment of the required non-orthogonality  condition{{\cite{das2012}}}, thus providing a testing ground for the theory of VIC {{\cite{agarwal_book}}}. For instance, as we schematically show 
in Fig.1, let us choose a pair of  ro-vibrational excited states $\mid 1 \rangle$ and $\mid 2 \rangle$ of a molecule with same rotational ($J_1 = J_2$) but two different vibrational ($ v_1 \ne v_2$) quantum numbers. Then obviously, the two  dipole moments for electric dipole transitions from these states to a common lower state are non-orthogonal. In this context, it is worth mentioning that the similar idea of non-orthogonality between two molecular dipole transitions has been suggested in the  context of incoherent pumping \cite{tscherbul2014}. The typical vibrational energy separation in low lying molecular states is in infra-red domain of the electromagnetic spectrum. In photoassociation (PA) spectroscopy of cold atoms, highly excited vibrational states with energy spacing lying in the microwave domain can be populated. Since microwave frequency is much greater than typical linewidth (a few MHz) of the molecular excited states accessible by PA, the second condition for VIC regarding energy spacing between the states will not be fulfilled. 
This difficulty can be circumvented by dressing the two vibrational states with a micro-wave (MW) field inducing magnetic M1 transitions. As a result, two dressed states $\mid e_1 \rangle$ and $\mid e_2\rangle$ which are the coherent superpositions between $\mid 1 \rangle$ and $\mid 2 \rangle$ will be formed. The spacing between these dressed states is Rabi frequency $\Omega$ if the MW field is tuned on resonance. Now, if the intensity of the MW field is such that $\Omega \simeq \sqrt{\gamma_1 \gamma_2}$ where $\gamma_{1(2)}$ is the spontaneous linewidth of $\mid 1\rangle$ $\left(\mid 2 \rangle\right)$ then the two decay pathways from the dressed levels are likely to interfere giving rise to VIC.    

The predicted bound state of two ultracold atoms in continuum is basically a multichannel resonance state with zero width. Ideally, this is perfect BIC in the absence of spontaneous emission.  Our results show that, even a vanishingly small spontaneous emission is sufficient to completely wash away any signature of BIC formation. Nonetheless, here we show that, the BIC can be absolutely protected against such highly destructive spontaneous emission by fulfilling the condition of VIC as we have just discussed above. We further show that if the VIC condition deviates even by 0.1 percent, the BIC can be destroyed completely.           

The paper is organized in the following way. In the next \cref{secmodel},  we present our model. The Hamiltonian describes interacting molecular bound and collisional continuum states. Our model takes into account the interaction of the excited 
molecular bound states with the vacuum of background electromagnetic field continuum to mimic the effects of spontaneous emission.  We then derive the analytical solution of the model and the condition for VIC in \cref{secsolution}. 
We use Feshbach projection operator method to eliminate the collisional and vacuum continua and thus to obtain a complex  effective Hamiltonian describing three interacting bound states.  The eigenvalues of the effective Hamiltonian are in general complex. A real eigenvalue of the Hamiltonian implies the formation of a BIC. Our analytical results show that, unless VIC condition is fulfilled, the Hamiltonian does not yield any real eigenvalue. In \cref{secpa} we discuss the possibility of detection of BIC by photoassociative spectroscopy. We present and  analyze our numerical results in \cref{rd}. The paper is concluded in \cref{secconcl}.

\section{The Model}
\label{secmodel}
We consider a magnetic Feshbach resonance between two ground-state (S + S) atoms under two-channel approximation \cite{mies2000,chin2010}. The Feshbach-resonant state as well as the continuum of scattering states can be coupled to molecular bound states in an excited potential by laser light {\cite{bauer2009,rempe1}}. In the excited manifold, we consider two vibrational states $\mid v_1\rangle$ and $\mid v_2\rangle$ with same rotational quantum number $J_1=J_2=J$.
These two excited states can be coupled by a microwave field due to magnetic dipole (M1) transition. The dressed states formed due to such coupling are
\begin{align}
\label{micro}
\mid e_1\rangle&=\cos\theta\mid v_1\rangle+\sin\theta\mid v_2\rangle\\
\mid e_2\rangle&=-\sin\theta\mid v_1\rangle+\cos\theta\mid v_2\rangle 
\end{align}
where $\tan 2\theta=\Omega_m/{\Delta_m}$, $\Omega_m=-\frac{1}{\hbar}\langle v_1J_1\mid \vec{\mu}.{\cal \vec{B}}\mid v_2 J_2\rangle$ is the microwave Rabi coupling with ${\Delta_m}=\omega_{12}-\omega_m$ being detuning of the microwave field of frequency $\omega_m$ from the vibrational spacing $\omega_{12}=\omega_{v_2}-\omega_{v_1}$, where $\hbar \omega_{v_i}$ is the eigen energy of the state $\mid v_i\rangle$. The purpose of this microwave coupling is to control the energy gap between the dressed states $\mid e_1\rangle$ and $\mid e_2\rangle$. The intensity of the microwave should be set at a level such that $\Omega_m \simeq \sqrt{\gamma_1 \gamma_2}$ where $\gamma_1$ and $\gamma_2$ are the two spontaneous emission linewidths. For homonuclear cold molecules composed of alkali atoms, the spontaneous linewidth of
the excited states is typically $\gamma/(2\pi) \simeq 5 $ MHz. For hetero-nuclear or homonuclear cold molecule composed of alkaline earth-type atoms the linewidth may be much smaller. The intensity of microwave dressing field should be set at a value  such that  $\Omega_m$ is comparable to $2\pi\gamma$, since near resonance for the microwave transition the two dressed states will be separated by about $\Omega_m$. Here we are dressing two excited states. Usually, a two-level system consisting of a ground-state (lossless) and an excited state with a small spontaneous emission linewidth is dressed by Rabi coupling the two states with a strong field. The dressing is effective if $\Omega_m$ exceeds $\gamma$. Now, in our case if the microwave field is too strong, then $\Omega_m$ would be too large and so would be spacing between $\mid e_1\rangle$ and $\mid e_2\rangle$. Therefore, the two excited states should be optimally Rabi coupled so that $\Omega_m \simeq \gamma$. Under such condition, the two decay paths will interfere destructively leading to the cancellation or inhibition  of the spontaneous emission. As we will discuss later, this condition is important for the achievement of BIC. Fig.\ref{model2} shows that the bare continuum of scattering states $\mid E\rangle_{b}$ in the open channel as well as the bound state $\mid c\rangle$ are coupled to the two dressed states $\mid e_1\rangle$ and $\mid e_2\rangle$ by two lasers $L_1$ and $L_2$, respectively.
\begin{figure}
 \includegraphics[width=0.9\linewidth]{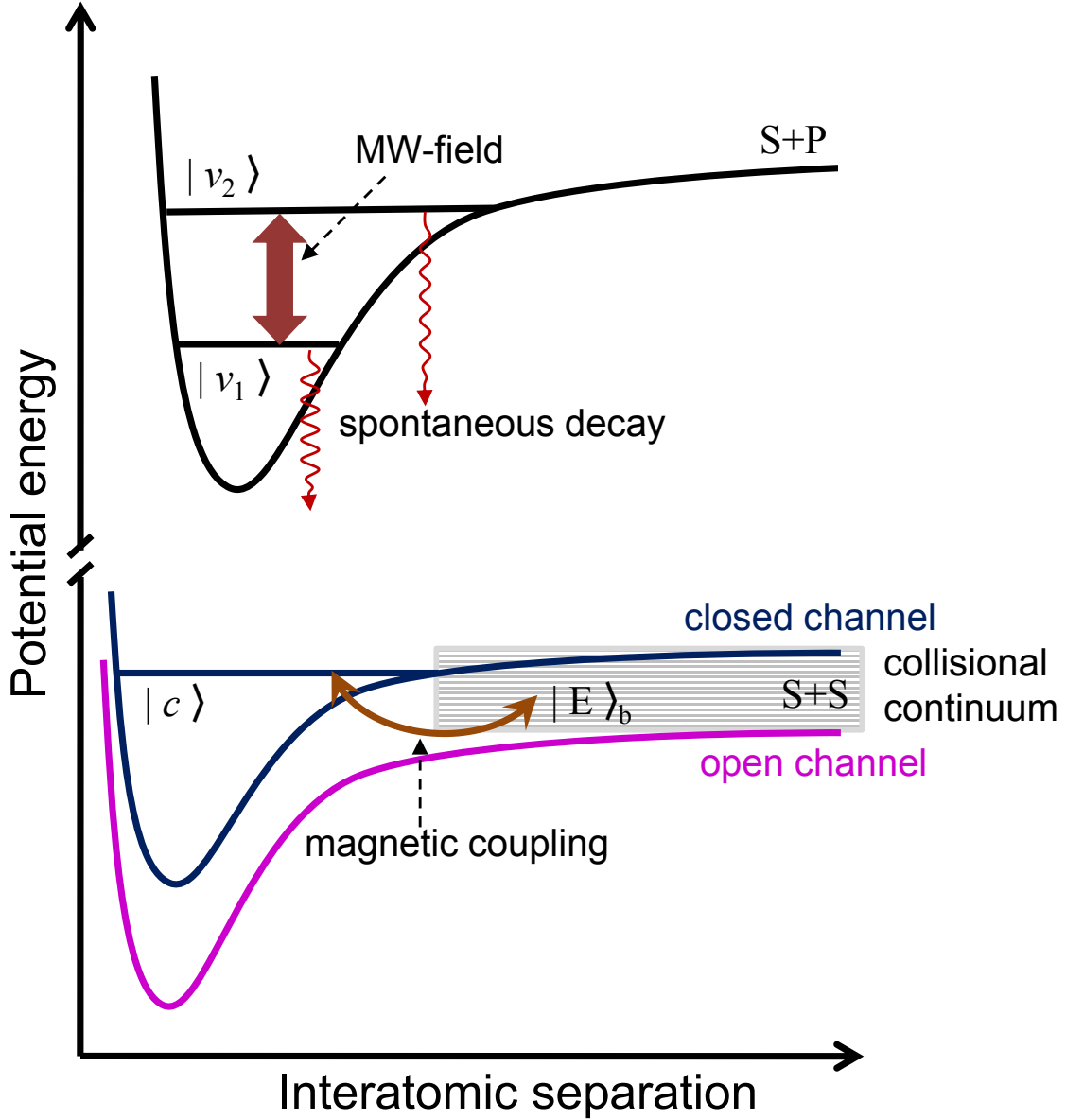}
\caption{ A schematic bare state picture of the system. In the two-channel approximation of a Feshbach resonance, the two ground-state potentials corresponding to the open and closed channels are shown. Also shown is an excited-state potential.
$\mid c\rangle$ is a quasi-bound state in the closed channel which is magnetically coupled to the scattering state in the open channel. $\mid v_1\rangle$ and $\mid v_2\rangle$ are excited bound states with vibrational quantum numbers $v_1$ and $v_2$, respectively; with the same rotational quantum number $J_1=J_2=J$. These two excited states are coupled through a microwave field, leading to the dressed excited states as shown in Fig.\ref{model2}}
\label{model}
\end{figure}

Next, we consider the following product states of the atomic kets and photon number kets 
\begin{align}
\mid3\rangle\equiv\mid c\rangle\otimes\mid N_1\omega_1, N_2\omega_2\rangle\equiv\mid c, N_1, N_2\rangle\nonumber 
\end{align}
\begin{align}
\mid2\rangle\equiv\mid e_2\rangle\otimes\mid N_1\omega_1, (N_2-1)\omega_2\rangle\equiv\mid e_2, N_1, N_2-1\rangle\nonumber 
\end{align}
\begin{align}
\mid1\rangle\equiv\mid e_1\rangle\otimes\mid (N_1-1)\omega_1,N_2\omega_2\rangle\equiv\mid e_1, N_1-1, N_2\rangle\nonumber 
\end{align}
\begin{align}
\mid E\rangle\equiv\mid E\rangle_b\otimes\mid N_1\omega_1,N_2\omega_2\rangle\equiv\mid E, N_1, N_2\rangle\nonumber 
\end{align}
as the basis of our analysis, where $N_1$, $N_2$ represent the photon numbers in the two lasers $L_1$, $L_2$, respectively. In absence of any spontaneous radiative decay, the Hamiltonian of the system can be expressed as $H=H^0_1 + V_1$, where
\begin{align}
\label{h01}
H^0_1&=\sum\limits_{N_1,N_2}\bigg[\Big(E_1+\left(N_1-1\right)\hbar\omega_1+N_2\hbar\omega_2\Big)\mid e_1, N_1-1, N_2\rangle\langle e_1, N_1-1, N_2\mid\nonumber\\&+\Big(E_2+N_1\hbar\omega_1+\left(N_2-1\right)\hbar\omega_2\Big)\mid e_2, N_1, N_2-1\rangle\langle e_2, N_1, N_2-1\mid\nonumber\\
&+\Big(E_3+N_1\hbar\omega_1+N_2\hbar\omega_2\Big)\mid c, N_1, N_2\rangle\langle c, N_1, N_2\mid\nonumber\\
&+\int\limits_{E_{th}}^\infty dE\Big(E+N_1\hbar\omega_1+N_2\hbar\omega_2\Big)\mid E, N_1, N_2\rangle\langle E, N_1, N_2\mid\bigg]
\end{align}
and
\begin{align}
\label{v1}
V_1=\sum\limits_{N_1,N_2}&\bigg[\hbar\Omega_{13} \mid e_1, N_1-1, N_2\rangle\langle c, N_1, N_2\mid+\hbar\Omega_{23} \mid e_2, N_1, N_2-1\rangle\langle c, N_1, N_2\mid \nonumber\\&+ \int\limits_{E_{th}}^\infty dE\Lambda_{1E}\mid e_1, N_1-1, N_2\rangle\langle E, N_1, N_2\mid \nonumber\\
+\int\limits_{E_{th}}^\infty& dE\Lambda_{2E}\mid e_2, N_1, N_2-1\rangle\langle E, N_1, N_2\mid  + \int\limits_{E_{th}}^\infty dE V_{3E}\mid c, N_1, N_2\rangle\langle E, N_1, N_2\mid\bigg] + h.c
\end{align}
Here, $E_j$ is the energy of the state $\mid e_j\rangle (j=1, 2)$ and $E_3$ is the energy of  $\mid c\rangle$. $\omega_j(j=1,2)$ denotes the frequency of the $L_j$ laser. $\Omega_{13}$ $\left(\Omega_{23}\right)$ represents the bound-bound  coupling between $\mid3\rangle$ and $\mid1\rangle$ $\left(\mid2\rangle\right)$. $\Lambda_{1E}$ and $\Lambda_{2E}$ are the free-bound coupling parameters, $V_{3E}$ is the coupling between $\mid3\rangle$ and $\mid E\rangle$ due to spin-dependent interactions. Explicitly, $\Omega_{i3} = (1/2) G_{i3} \sqrt{N_i +1}$, $(i=1,2)$ where $G_{i3} = - (1/\hbar) \langle i \mid {\mathbf D}_i \cdot {\mathbf E}_i \mid 3 \rangle$ with ${\mathbf E}_i$ being the laser field amplitude corresponding to the zero photon;  and ${\mathbf D}_i$ being the molecular transition dipole moment between states $\mid e_i \rangle$ and $\mid c \rangle$.
\begin{figure}
\begin{center}
\includegraphics[width=0.8\linewidth]{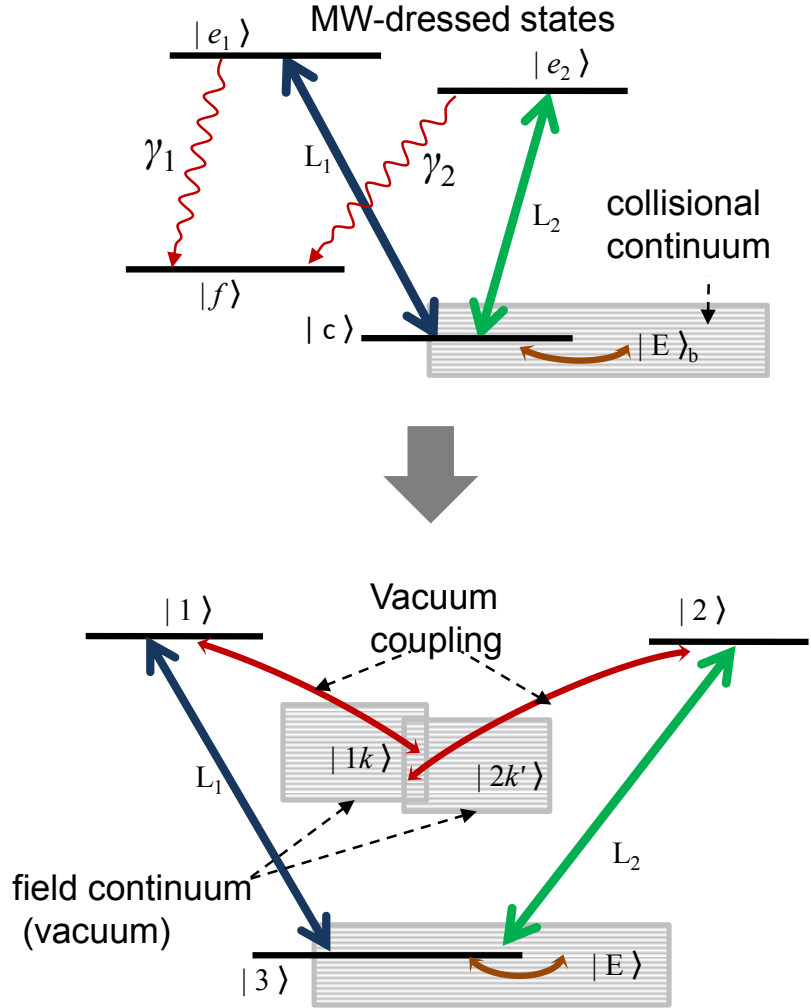}
\caption{The states $\mid e_1\rangle$ and $\mid e_2\rangle$ are the microwave dressed states which can decay spontaneously to $\mid f\rangle$ with rate $\gamma_{1}$ and $\gamma_{2}$, respectively. The two lasers $L_1$ and $L_2$ that drive PA transitions from the magnetic Feshbach (MF) resonant collisional state $\mid E\rangle$ to $\mid e_1\rangle$ and $\mid e_2\rangle$, respectively. The atom-photon composite state representation (see the text) where $\mid 1\rangle$ and $\mid 2\rangle$ interact with the continua $\mid 1k\rangle$ and $\mid 2k^{^\prime}\rangle$ of background electromagnetic fields through coupling via vacuum fields. Here, under certain conditions, the two decay paths leading to $\mid f\rangle$ can cancel each other, nullifying the spontaneous emission.}
\label{model2}
\end{center}
\end{figure}

Now, we consider spontaneous radiative decay from $\mid e_1\rangle$ and $\mid e_2\rangle$. Usually, these states decay to nearly all continuum and bound states of the open and closed channels. However, since both MW-dressed these states have the same molecular angular momentum and  components of the two vibrational states, it is expected that both these states will be coupled to one or multiple final states by vacuum fields. As a model system, we assume that $\mid e_1\rangle$ and $\mid e_2\rangle$ predominantly decay to a common external state $\mid f\rangle$ as shown in Fig.\ref{model2} and neglect all other spontaneous decay pathways. We describe these two decay processes by introducing the following photon continuum channels \cite{haan1987}
\begin{align}
\label{1k}
\mid 1k\rangle\equiv\mid f\rangle\otimes \mid N_1\omega_1, N_2\omega_2, k\rangle\equiv\mid f, N_1, N_2, k\rangle
\end{align}
\begin{align}
\label{2kp}
\mid 2k^{^\prime}\rangle\equiv\mid f\rangle\otimes \mid N^{^\prime}_1\omega_1, N^{^\prime}_2\omega_2, k^{^\prime}\rangle\equiv\mid f N^{^\prime}_1, N^{^\prime}_2, k^{^\prime}\rangle
\end{align}
where $k$ represents the energy of the spontaneously emitted photon.
The two radiative decay channels $\mid 1k\rangle$ and $\mid 2k^{^\prime}\rangle$ are normalized in the following way
\begin{align}
%\label{alphabeta}
\langle 1k\mid 2k^{^\prime}\rangle=\delta_{N_1,N^{^\prime}_1} \delta_{N_2,N^{^\prime}_2}\delta\left(k-k^{^\prime}\right)
\end{align}
When spontaneous decay is incorporated, the Hamiltonian becomes $H = H^0 + V$ with $H^0 = H^0_1 + H^0_2$ and $V = V_1 + V_2$ where

\begin{align}
\label{h02}
H^0_2=\sum\limits_{N_1,N_2}\int\limits_0^\infty kdk\mid 1k\rangle\langle 1k\mid+\sum\limits_{N^{^\prime}_1,N^{^\prime}_2}\int\limits_0^\infty k^{^\prime}dk^{^\prime}\mid 2k^{^\prime}\rangle\langle 2k^{^\prime}\mid
\end{align}
\begin{align}
\label{v2}
V_2=\sum\limits_{N_1,N_2}\int\limits_0^\infty dkV_{1k}\mid1\rangle\langle 1k\mid+\sum\limits_{N^{^\prime}_1,N^{^\prime}_2}\int\limits_0^\infty dk^{^\prime}V_{2k^{^\prime}}\mid2\rangle\langle 2k^{^\prime}\mid+h.c
\end{align}
and the basis are expanded to ${\Big[}\mid 1\rangle,\mid 2\rangle,\mid 3\rangle,\mid E\rangle,\mid 1k\rangle,\mid 2k^{^\prime}\rangle{\Big]}$. We treat laser fields semiclassically which is equivalent to a quantum treatment if the fields are in coherent states with the average photon numbers being large. This was discussed extensively especially by C. Cohen-Tannoudji \cite{cohen1977,cohen_book} who examined quantized dressed state and how the semiclassical description emerges in the limit of large photon numbers.

Next, following the Feshbach projection operator technique, we eliminate all the continuum states and find the effective Hamiltonian $H_{eff}$ in the atomic subspace ${\Big[}\mid e_1\rangle,\mid e_2\rangle,\mid c\rangle{\Big]}$. $H_{eff}$ is non-hermitian. The imaginary part of the Hamiltonian describes interaction of the subspace with the rest of the complete space. A BIC refers to an eigenstate of this $H_{eff}$ having real eigenvalue.
In the following, we write all the matrix elements $H_{eff}$ of the effective Hamiltonian following the derivations done in \cref{ap1}. 
\begin{subequations}
\begin{align}
\label{h11}
&\langle e_n\mid H_{eff}\mid e_n\rangle=\left(E_n - \hbar\omega_n + E^{^{\rm{sh}}}_{n}\right) - \frac{i\hbar}{2}\left(\Gamma_{n}+\gamma_{n}\right), \hspace{1cm} n=1,2\\
&\langle c\mid H_{eff}\mid c\rangle=E_3+E^{^{\rm{sh}}}_{F}-\frac{i\hbar}{2}\Gamma_{F}\\
\label{h12}
&\langle e_1\mid H_{eff}\mid e_2\rangle=\alpha-\frac{i\hbar}{2}\left[\gamma_{_{LIC}}+\gamma_{_{VIC}}\right]\\
\label{hn3}
&\langle e_n\mid H_{eff}\mid c\rangle=\beta_n+\hbar\Omega^{^\prime}_{nc}-\frac{i\hbar}{2}\Gamma_{nF}, \hspace{1cm} n=1,2\\
\label{h3n}
&\langle c\mid H_{eff}\mid e_n\rangle=\langle e_n\mid H_{eff}\mid c\rangle
\end{align}
\end{subequations}
Here, $E^{^{\rm{sh}}}_{1}$ $\Big(E^{^{\rm{sh}}}_{2}\Big)$ is the laser-induced shift of the state $\mid e_1\rangle$ $\Big(\hspace{1mm}\mid e_2\rangle\hspace{1mm}\Big)$ and $E^{^{\rm{sh}}}_{F}$ is magnetic field induced shift of the state $\mid c\rangle$ which are shown explicitly in \cref{ap1}. $\gamma_{_{LIC}}$ and $\gamma_{_{VIC}}$ are the terms that arise respectively due to the laser-induced and vacuum-induced coherence with the latter existing only in case of non-orthogonal transition dipole moments. We assume all the vacuum induced shifts to be negligible. The shifted energy $E_3+E^{^{\rm{sh}}}_{F}$ of the state $\mid c\rangle$ can be related to the magnetic field dependent scattering length $a_s({\cal B})$ of the colliding atom-pairs in the limit $E_3\to 0$ as \cite{deb2014}
\begin{align}
E_3+E^{^{\rm{sh}}}_{F}=-\left(k_ca_s\right)^{-1}\frac{\hbar\Gamma_F}{2}
\end{align}
where $k_c$ is the wave number related to the collisional energy $E={\hbar^2k_c^2}/{2\mu}$, $\mu$ being the reduced mass of the atom pairs and $\Gamma_F$ is the width of FBR. 
$\beta_{n}$ and $\alpha$ are defined as the following principal value integrals consist of laser and magnetic coupling parameters
\begin{align}
\beta_n={\cal P}\int\limits_0^\infty dE \frac{\Lambda^{^\prime}_{nE}V^{^{\prime *}}_{3E}}{E_3-E}
\end{align}
\begin{align}
\alpha&={\cal P}\int\limits_0^\infty dE \frac{\Lambda^{^\prime}_{1E}\Lambda^{^{\prime *}}_{2E}}{E_3-E}
\end{align}
$\Lambda^{^\prime}_{nE}$, $\Omega^{^\prime}_{nE}$ and $V^{^\prime}_{3E}$ are the coupling parameters in the regime of classical field approximation as mentioned in \cref{ap1}.
For notational convenience we introduce dimensionless parameters
$\delta_{n}=\left(E_1-\hbar\omega_{L_n}+E^{^{\rm{sh}}}_{n}\right)/\left(\hbar \Gamma_F/2\right),\hspace{4mm}$ $\delta=\alpha/\left(\Gamma_F/2\right),\hspace{4mm}$ 
$g_n=\Gamma_{n}/\Gamma_F$,$\hspace{4mm}g_{12}=\gamma_{_{LIC}}/\Gamma_F$,$\hspace{4mm} {\eta}=\gamma_{_{VIC}}/\Gamma_F \hspace{2mm}$and$\hspace{2mm}      
\tilde{\gamma}_n=\gamma_{n}/\Gamma_F\hspace{2mm}$for$\hspace{2mm} n=1,2$. 
We also introduce the Fano-Feshbach asymmetry parameters $q_{n} = \left({\beta + \hbar\Omega^{^\prime}_{n3}}\right)/{\left(\hbar\Gamma_{nF}/2\right)}$ where $n=1,2$.
In terms of these parameters we write $H_{eff}$ in matrix form as follows.
\begin{align}
 H^{eff}=\frac{\hbar\Gamma_F}{2} [A + iB]
\end{align}
\begin{align}
\mathbf{A} = \left(
\begin{array}{c c c}
{\delta}_{1} & \delta  & q_{1}\sqrt{g_1} \\
 \delta & {\delta}_{2} & q_{2}\sqrt{g_2} \\
q_{1}\sqrt{g_1} & q_{2}\sqrt{g_2} & - (k_ca_s)^{-1}  
\end{array}
\right) \nonumber\\ \nonumber\\
\end{align}
\begin{align}
 \mathbf{B} = - \left(
\begin{array}{c c c}
g_1+{\tilde\gamma}_1 &  g_{12}+\eta  & \sqrt{g_1} \\
g_{12}+\eta & g_2+{\tilde\gamma}_2 & \sqrt{g_2} \\
\sqrt{g_1} & \sqrt{g_2} & 1 
\end{array}
\right)  \nonumber\\ \nonumber\\
\end{align}

\section{Solution: The effect of VIC}
\label{secsolution}
A BIC is characterized by an eigenstate of $H_{eff}$ with real eigenvalue. The effective Hamiltonian may have one or more real eigenvalues when the eigenvectors of {\bf B} with zero eigenvalues become simultaneous eigenvectors of {\bf A} with real eigenvalues. First we determine the condition for at least one zero eigenvalue of {\bf B}. The secular equation for the {\bf B} matrix is 
\begin{align}
\label{bsec}
x^3 + G_2 x^2 + G_1 x + G_0 = 0
\end{align}
where
\begin{align}
G_2=&1 + g_1 + g_2 + {\tilde\gamma}_1 + {\tilde\gamma}_2\\
G_1=&{\tilde\gamma}_1+{\tilde\gamma}_2+g_1{\tilde\gamma}_2+g_2{\tilde\gamma}_1+{\tilde\gamma}_1{\tilde\gamma}_2-\eta^2-2\eta g_{12}\nonumber\\
G_0 = &-\eta^2+\tilde\gamma_1{\tilde\gamma}_2
\end{align}
In the absence of spontaneous radiative decay i.e $\tilde\gamma_1=0$, $\tilde\gamma_2=0$ and $\eta=0$, we reproduce the same secular equation which was derived in a previous work \cite{deb2014}. It is evident from Eq. (\ref{bsec}) that $\bf B$ can have one zero eigenvalue only when $G_0$ vanishes which gives
\begin{align}
\label{eta2}
 \eta^2-\tilde\gamma_1{\tilde\gamma}_2=0
\end{align}
Thus,
\begin{align}
\label{b0}
\eta=\pm\sqrt{\tilde\gamma_1{\tilde\gamma}_2}
\end{align}
We consider only one root $\eta=\sqrt{\tilde\gamma_1{\tilde\gamma}_2}$, which may be interpreted as the vacuum coherence induced decay between the two excited states. The corresponding eigenvector is given by
\begin{align}
\label{bic}
 {X} = C \left(
\begin{array}{c }
    x_1  \\
    x_2\\
    1
\end{array}
\right ) 
\end{align} 
where
\begin{align}
\label{x2}
x_{1}=\frac{\sqrt{\tilde\gamma_2}}{\sqrt{g_2\tilde\gamma_1}-\sqrt{g_1\tilde\gamma_2}}
\end{align}
and
\begin{align}
\label{x3}
x_2=\frac{-\sqrt{\tilde\gamma_1}}{\sqrt{g_2\tilde\gamma_1}-\sqrt{g_1\tilde\gamma_2}}
\end{align}
$C=\left(x_1^2+x_2^2+1\right)^{\frac{1}{2}}$ is the normalization constant. The state given by (\ref{bic}) will be a BIC if it becomes an eigenvector of $\bf A$ with real eigenvalue $\lambda$.
\begin{align}
{\bf A}X=\lambda X
\end{align}
Consequently, we get the following three equations to be satisfied
\begin{align}
\label{al1}
{\delta}_{1}-\lambda-g_1q_1+\sqrt{\frac{\tilde\gamma_1}{\tilde\gamma_2}}\left(g_{12}q_1-\delta\right)=0
\end{align}
\begin{align}
\label{al2}
\delta-g_{12}q_2+\sqrt{\frac{\tilde\gamma_1}{\tilde\gamma_2}}\left(g_2q_2-\delta_2+\lambda\right)=0
\end{align}
\begin{align}
\label{al3}
g_1\left(q_1+(k_ca_s)^{-1}+\lambda\right)-g_{12}\sqrt{\frac{\tilde\gamma_1}{\tilde\gamma_2}}\left(q_2+(k_ca_s)^{-1}+\lambda\right)=0
\end{align}
We can assume $(k_ca_s)^{-1}\rightarrow 0$ if the magnetic field is tuned closed to Feshbach resonance. The quantities ${\tilde \gamma}_1$, ${\tilde \gamma}_2$, $q_1$ and $q_2$ pertain to a particular system under consideration. For a particular set of values of these quantities together with the values of $g_1$, $g_2$ and $\delta$, Eqs. (\ref{al1}), (\ref{al2}) and (\ref{al3}) can be solved for $\delta_1$, $\delta_2$ and $\lambda$.

It is important to note that in the absence of VIC, i.e $\eta=0$, the secular equation Eq. (\ref{bsec}) of matrix $\bf B$ cannot have a zero eigenvalue due to the presence of spontaneous emission as $G_0$ remains non-zero. Thus, the presence of spontaneous emission alone, rules out the possibility of a BIC. Only when VIC is present and nullifies the effect of spontaneous emission through Eq. (\ref{eta2}) making $G_0=0$, a BIC is obtained.

The existence of the VIC parameter $\eta$ crucially depends on the non-orthogonality of the transition dipole moments associated with the spontaneous emissions. In case of molecular bound states, two different vibrational levels with same rotational quantum number facilitate such non-orthogonality. In \cref{rd}, we show that in the most common realistic situations where the spontaneous decay is present without any VIC, a BIC cannot be found. Only when the VIC term is non-zero and Eq. (\ref{eta2}) is fulfilled, a BIC can be created.

It is also important to note that an effective destructive interference between the two spontaneous emission pathways will occur if the frequency spacing between two decaying states is much less than $\eta$. To ensure this condition, we consider microwave dressed states $\mid e_1\rangle$ and $\mid e_2\rangle$ whose energy difference can be tuned by the microwave dressing field to the desired level.

\section{Photoassociative detection of BIC {Via vanishing width line in Fano spectrum}}
\label{secpa}

The probability of the Photoassociative transition $\mid E\rangle_b\rightarrow \mid e_n\rangle (n=1,2)$ is modified due to the formation of BIC and it can be expressed as 
\begin{align}
 P_n= \int dE \mid\langle e_n\mid E+\rangle\mid^2
\end{align}
where dressed continuum state $\mid E+\rangle=\Omega_+\mid E\rangle_b$ and $\Omega_+$
denotes a M\o{}ller operator \cite{deb2014}. Here, the quantity 
\begin{align}
 S_n(E)=\mid\langle e_n\mid E+\rangle\mid^2
\end{align}
is the photoassociation probability per unit collision energy. 
Now,
\begin{align}
\langle e_n\mid E+\rangle &=\langle e_n\mid (E-H_{eff})^{-1}V\mid E\rangle_b
\end{align}
Using the standard form for the inverse of a matrix, we find
\begin{align}
\langle e_n\mid \Omega_+\mid E\rangle_b=\sqrt{\frac{2}{\pi\hbar\Gamma_F}} 
\frac{1}{{\rm Det}\left[\tilde{E}-\tilde{H}^{^\prime}_{eff}\right]}\times
\left[ {\mathscr A}_{n 1 } \sqrt{g_1}+{\mathscr A}_{n 2 } \sqrt{g_2} + {\mathscr A}_{n 3} \right]
\label{eqn}
\end{align}
where \hspace{2mm}$\tilde{E}=E/E_F $,\hspace{2mm}$E_F=\hbar\Gamma_F/2$,\hspace{2mm} $\tilde{H}^{^\prime}_{eff} = \mathbf{A} + i \mathbf{B}$ and 
${\mathscr A}$ is the transpose of the co-factor matrix of $(\tilde{E} - \tilde{H}_{eff} )$.
From Eq. (\ref{eqn}), the denominator part 
\begin{align}
{\rm Det}[\tilde E -\tilde H_{eff}]=\prod^3_{i=1} (\tilde E -\tilde E_{i})
\end{align}
where $\tilde E_i$ denotes an energy eigenvalue of $\tilde H_{eff}$, may become zero for a real $\tilde E_i$ only including the numerator part is also zero for that $\tilde E_i$. The system exhibits an exceptionally sharp Fano-like spectrum {\mbox{\cite{fano1,fano2}}} when the imaginary part of a complex eigenvalue of $H_{eff}$ is extremely small i.e one of the eigenvalues of $H_{eff}$ become real.

\section{Results and discussions}
\label{rd}

\begin{figure}
\begin{center}
\vspace{1.5cm}
\psfrag{a}[b][b][1.0]{${\cal S}_1\left(\tilde E\right)$}
\psfrag{b}[b][b][1.0]{$\tilde E$}
\includegraphics[width=0.8\linewidth]{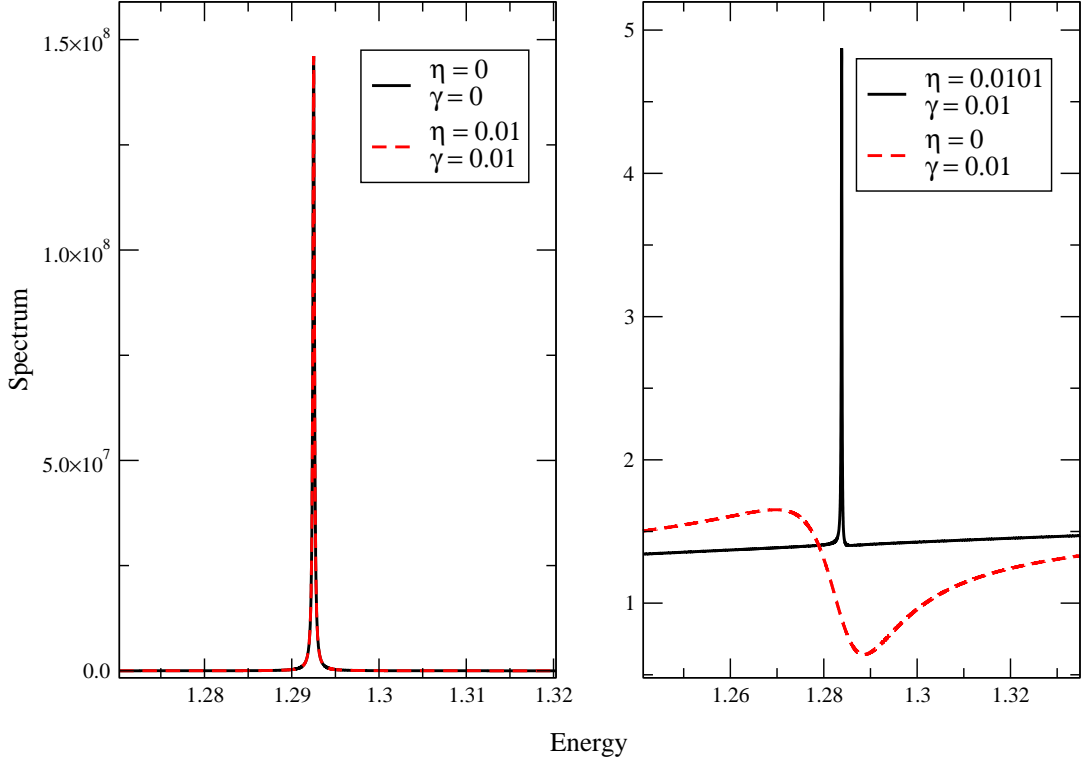}
\caption{{Dimensionless spectrum ${\cal S}_1$($\tilde E$) as a function of scaled energy $\tilde E$. The common parameters for the above four spectra: $g_1=4,$ $g_2=1.91,$ whereas $g_2=2$ for BIC, $q_1=-0.8,$ $ q_{2}=-0.6,$ $\delta_1=0.45,$ $\delta_2=1.88$ and $\delta=0.1$. The black, solid spectrum in the left panel indicates the formation of BIC in absence of spontaneous emission ${\tilde\gamma}_1={\tilde\gamma}_2=\gamma=0$ corresponding to the eigenvalue $\tilde E_1=1.29-10^{-4}i$. The other two roots are $\tilde E_2=-0.538-6.459i$ and $\tilde E_3=1.571-0.450i$. The same spectrum is reproduced even in presence of spontaneous emissions when VIC condition is incorporated {(\ref{eta2})} which is shown by red, dashed curve. In the right hand panel, the black, solid line indicates that  
 BIC is restored with $\gamma=0.01$ when $\eta=0.0101$ although the height of the peak is decreased. The red dashed line indicates that presence of very small amount of spontaneous emission $\gamma=0.01$ destroys the BIC when the VIC parameter $\eta=0$. } }
\label{spectrum}
\end{center}
\end{figure}

{In Fig.{\ref{spectrum}}, we show the scaled photoassociative absorption spectra ${\cal S}_1\left(\tilde E\right)=S_1(E)E_F$ as a function of the scaled energy $\tilde E=E/E_F$ for a particular set of parameters as given in the caption. When the parameters are set to match a perfect BIC condition, no spectrum can be obtained as the width of the resonance becomes zero. For this reason, we slightly deviate the value of $g_1$ from the required value to satisfy BIC condition. This allows a very small imaginary part in the real root and show up an ultra-narrow resonance line.

In the first panel, with the given set of parameters, the exceedingly narrow feature of the spectrum indicates the presence of BIC corresponding to an eigenvalue $\tilde E=1.30-10^{-6}i$ in absence of any spontaneous decay. The BIC is sustained in presence of spontaneous emissions ${\tilde\gamma}_1={\tilde\gamma}_2=\gamma=0.01$, if we include the VIC term satisfying the condition {(\ref{eta2})}. The corresponding curves are hardly distinguishable as they almost exactly coincide. In the second panel, we show that if the condition {(\ref{eta2})} deviates slightly, it affects the BIC drastically with the height of the spectrum being severely decreased. The BIC is destroyed showing a widened peak when we switch off the VIC term. Therefore, VIC parameter has a decisive role in obtaining BIC in presence of spontaneous emissions}. 

\begin{figure}[htp]
\begin{center}
\psfrag{s}[b][b][1.0]{${\cal S}_1\left(\tilde E\right)$}
\psfrag{e}[b][b][1.0]{$\tilde E$}
\includegraphics[width=0.8\linewidth]{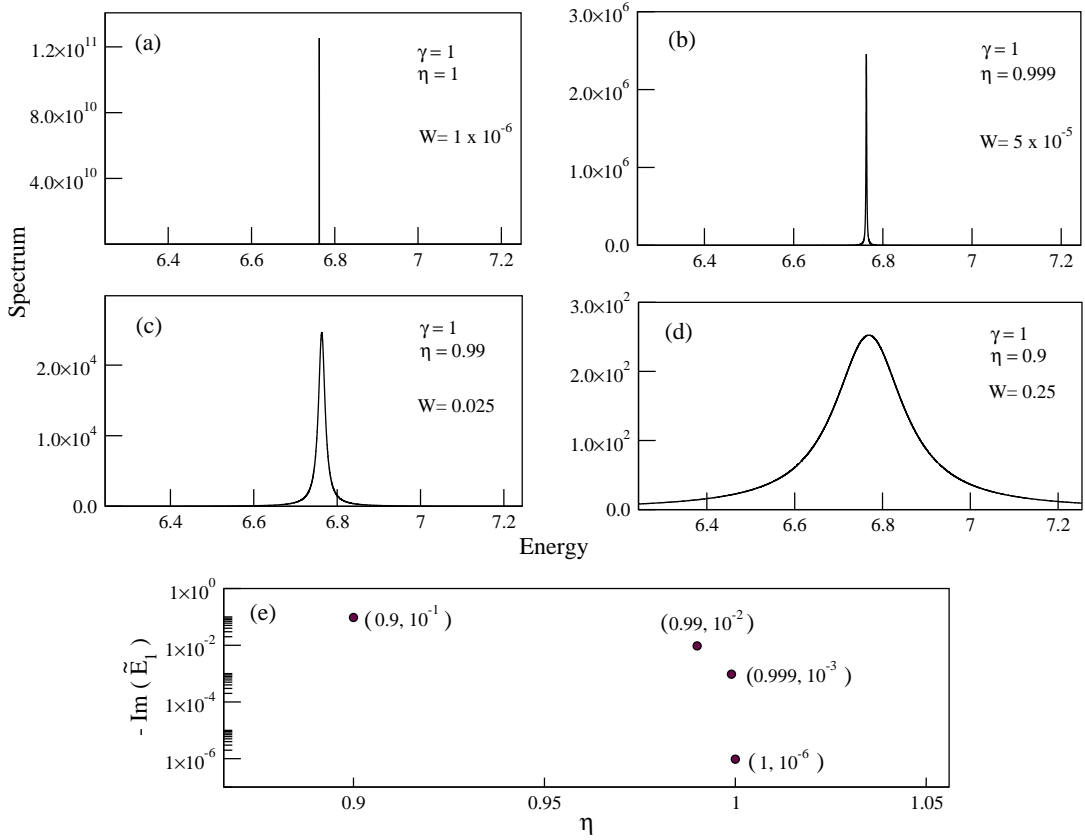}
\caption{{Dependence of the nature of spectrum showing the extreme sensitivity on the deviation of $\eta $ from that satisfying BIC condition of Eq.{(\ref{eta2})}. The parameters chosen are $\gamma_1=\gamma_2=\gamma=1$, $g_1=3.01,$ whereas $g_1=3$ for BIC. $g_2=2,$  $q_1=-0.8,$ $ q_{2}=0.54,$ $\delta_1=6.4,$ $\delta_2=6.6$ and $\delta=0.1$. The eigenvalue closest to that of a BIC is $\tilde E_1=6.763-10^{-6}i$. The other two eigenvalues are $\tilde E_2=6.051-7.182i$ and $\tilde E_3=0.227-8.828i$. The value of $\eta$ is gradually deviated as {\textbf {(a)}}: $\eta=1$, {\textbf {(b)}}: $\eta=0.999$ and {\textbf {(c)}}: $\eta=0.99$. {\textbf {(d)}}: $\eta=0.9$}. The spectral width $W$ is mentioned in the corresponding panels. {\textbf {(e)}}: For the above four values of $\eta$, the movement of the imaginary part of $\tilde E_1$ shown.}
\label{eta-spectrum}
\end{center}
\end{figure}
\begin{figure}
 \begin{center}
  \includegraphics[width=0.7\linewidth]{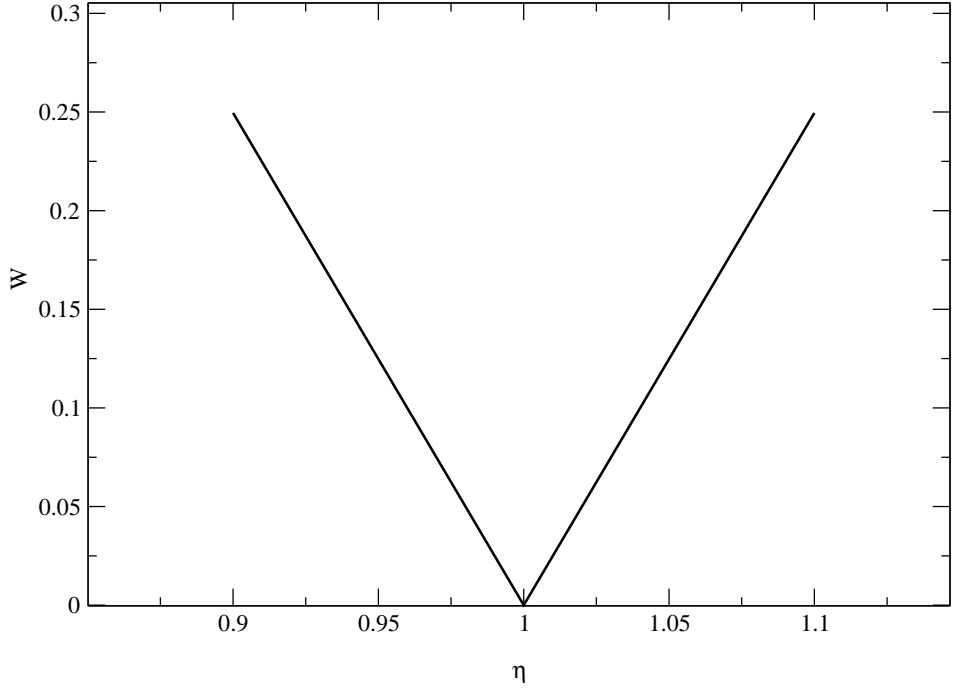}
    \caption{{The width (W) of the BIC peak is plotted as function of $\eta$ for a particular set of parameters: $g_1=3,$ $g_2=2,$  $q_1=-0.8,$ $ q_{2}=0.54,$ $\delta_1=6.4,$ $\delta_2=6.6$, $\delta=0.1$, and $\gamma=1$ }}
   \label{width}
 \end{center}
\end{figure}

 
The significance of VIC is shown in another way in Fig.{\ref{eta-spectrum}}. It appears that the ultra-narrow nature of the spectra is drastically broadened if we change the value of $\eta$ to even $1\%$ from that satisfying BIC condition {(\ref{eta2})}. The physical origin of such spectral behavior may lie in the strong dependence of the decoherence of the "Fano" coherence \cite{koyu2018} or excited molecular dark state \cite{saha2014} on the VIC. In particular, the decoherence has been shown to be highly sensitive to the variation of the alignment between the two dipole moments \cite{koyu2018}, or in other words, to the degree of non-orthogonality of the dipole moments. Here, we plot the spectrum considering another set of parameters given in the caption. Since $\gamma=1$, the BIC condition {(\ref{eta2})} is satisfied when $\eta=1$ and the eigenvalue is $\tilde E=6.76-10^{-6}i$. The width of the spectrum happens to be $W\simeq 1\times 10^{-6}$ which is shown in panel-(a). We take three values of $\eta$ that gradually deviates from unity and show how the nature of the spectrum changes so dramatically. From panel-(b) to (d) in Fig.{\ref{eta-spectrum}}, the value of $\eta$ is considered to be $0.999$, $0.99$ and $0.9$ for which the width become $2.0\times10^{-5}$, $0.025$ and $0.25$, respectively. The width of the peak is calculated as the difference between the two values of $\tilde E$ where the value of ${\cal S}_1\left({\tilde E}\right)$ decreases to $1/e$ of its maximum value. Subsequently, the height of the resonance peaks decrease by two or three orders of magnitude. In panel-(e), we show the movement of the imaginary part of $\tilde E_1$, as $\eta$ deviates from unity at which BIC appears. As $\eta$ deviates gradually form unity to $0.999$, $0.99$ and $0.9$, the imaginary part of $\tilde E_1$ changes to $-10^{-3}$, $-10^{-2}$ and $-10^{-1}$, respectively, leading to the drastic change in spectral width and height as shown in Fig.\ref{eta-spectrum}. In Fig.{\ref{width}}, we show 
the variation of width ($W$) as a function of $\eta$ in the close vicinity of $\eta=1$.

It is important to note that a BIC can serve as a tool for the efficient production of Feshbach molecules \cite{deb2014}. In general, a BIC is different from a Feshbach molecular state as it is in linear superposition of all the excited as well as the ground bound states and the eigenvalue may lie above the threshold of collisional continuum. But a Feshbach molecule is a negative-energy bound state lying below the threshold. Generally, a Feshbach molecule is produced by an inelastic collision involving three-body
process, but for the formation of a BIC, three-body process is not essential. It is evident from Eqs. (\ref{x2}) and (\ref{x3}), the more we offset the ratio $g_1/g_2$  from $\gamma_1/\gamma_2$,  the more does the superposition lean towards $\mid c\rangle$ which is a multichannel quasi-bound state. The wavefunction corresponding to a BIC expected to be very similar to that of the usual bound states having an exponential decay tail at large inter-atomic separation.

\section{Conclusions}
\label{secconcl}
In conclusion, we have demonstrated how to suppress the deleterious effects of spontaneous emission in creating diatomic bound state in continuum by manipulating the quantum states of a pair of interacting cold atoms with external fields. We have shown that the
suppression crucially depends on the use of appropriately tailor-made vacuum-induced coherence, the effectiveness of which essentially relies on the availability of two non-orthogonal transition dipole moments for two closely-spaced excited states. We emphasize that the non-orthogonality condition can be fulfilled by employing two molecular excited states having same rotational quantum number and differing only in vibrational quantum number. The energy spacing between these two states is required to be comparable to the geometric mean
of the two spontaneous emission linewidths. We have suggested that this requirement on energy spacing can be fulfilled by dressing the two excited states with a microwave field.
In passing, we remark that a variant of our model may be possible by considering the orbital Feshbach resonance between ground and metastable 
atoms \cite{kato2013,pagano2015,hofer2015} and coupling ground-state molecular bound states with the resonant states. 
Metastability of the excited states will give some advantage. We hope to address this issue 
in our future communication. Finally, we stress that with the recent development in the technology of coherent control over the states of cold atoms and cold molecules, the prospect for the realization of our proposed cold-atom BIC appears to be quite promising.
\section{Acknowledgment}
Dibyendu Sardar is grateful to CSIR, Government of India, for a support.

\color{black}
\appendix
\section{Derivation of the effective Hamiltonian}
\label{ap1}
Following \cref{secmodel}, in the classical field limit we can replace $\Omega_{i3}$ by  $\Omega_{i3}^{^\prime}= (1/2) G_{i3} \sqrt{\langle N_i \rangle }$, where $\left\langle N_i\right\rangle$ is the average photon number of $i-$th
laser. Similarly, $\Lambda_{iE}^{^\prime} \propto \sqrt{\langle N_i \rangle }$ while 
$ V^{^\prime}_{ik}=\left\langle V_{ik}\right\rangle$. However, the spontaneously emitted field is treated quantum mechanically as a one-photon state. So, the states $\mid1k\rangle$ and $\mid2k^{^\prime}\rangle$ defined by Eqs. (\ref{1k}) and (\ref{2kp}) reduce to $\mid f,k\rangle$ and $\mid f,k^{^\prime}\rangle$, respectively.
The semiclassical Hamiltonian in the rotating wave approximation is then given by
\begin{align}
H^{^\prime}=H^{^\prime}_0+V^{^\prime}
\end{align}
where
\begin{align}
\label{hp01}
H^{^\prime}_0&=(E_1-\hbar\omega_1)\mid e_1\rangle\langle e_1\mid+(E_2-\hbar\omega_2)\mid e_2\rangle\langle e_2\mid+E_3\mid c\rangle\langle c\mid \nonumber\\
+&\int\limits_{0}^\infty E \mid E\rangle_b\langle E\mid_b dE+ \int\limits_0^\infty kdk\mid f,k\rangle\langle f,k\mid+\int\limits_0^\infty k^{^\prime}d{k^{^\prime}}\mid f,k^{^\prime}\rangle\langle f,k^{^\prime}\mid
\end{align}
and
\begin{align}
\label{vp1}
V^{^\prime}_1=&\bigg[\hbar\Omega^{^\prime}_{13} \mid e_1\rangle\langle c\mid + \hbar\Omega^{^\prime}_{23} \mid e_2\rangle\langle c\mid + \int\limits_{0}^\infty \Lambda^{^\prime}_{1E}\mid e_1\rangle\langle E\mid dE+\int\limits_{0}^\infty \Lambda^{^\prime}_{2E}\mid e_2\rangle\langle E\mid dE\nonumber\\
& + \int\limits_{0}^\infty V^{^\prime}_{3E}\mid c\rangle\langle  E\mid dE+ \int\limits_0^\infty d{k}V^{^\prime}_{1k}\mid e_1\rangle\langle f,k\mid+\int\limits_0^\infty d{k^{^\prime}}V^{^\prime}_{2k^{^\prime}}\mid e_2\rangle\langle f,k^{^\prime}\mid\bigg] + h.c
\end{align}

Next, we define the projection operators for our model as
\begin{align}
\label{p}
P&=\mid e_1\rangle\langle e_1\mid +\mid e_2\rangle\langle e_2\mid + \mid c\rangle\langle c\mid
\end{align}
\begin{align}
\label{q}
Q&=\int\limits_0^\infty\mid E\rangle_b \langle E\mid_b dE + \int\limits_0^\infty d{k}\mid f,k\rangle\langle f,k\mid+\int\limits_0^\infty d{k^{^\prime}}\mid f,k^{^\prime}\rangle\langle f,k^{^\prime}\mid
\end{align}
where $P$ and $Q$ satisfy the conditions
$PP=P$, $QQ=Q$, $PQ=QP=0$ and $P+Q=1$. For the reduced Hamiltonian $H^{^\prime}$ obtained above, the resolvent operators $G(z)= (z-H^{^\prime})^{-1}$ and $G_0(z)= (z-H^{^\prime}_0)^{-1}$ are related as 
\begin{align}
\label{g}
G=G_0+G_0V^{^\prime}G
\end{align}
and
\begin{align}
PG(z)P=(z-PH^{^\prime}_0P-PRP)^{-1}
\end{align}
where
\begin{align}
\label{r}
R=V^{^\prime}+V^{^\prime}Q\frac{1}{z-QH^{^\prime}Q}QV^{^\prime}
\end{align}
where $z$ is a variable in the complex energy plane. The effective Hamiltonian in subspace defined by $P$ is given by  
\begin{align}
\label{heff1}
H_{eff} = PH^{^\prime}_0P + PRP
\end{align}
Combining (\ref{r}) and (\ref{heff1}) we write
\begin{align}
\label{heff2}
H_{eff}&=PH^{^\prime}_0P+PRP\nonumber\\
&=PH^{^\prime}_0P+PV^{^\prime}P+PV^{^\prime}Q\frac{1}{z-QH^{^\prime}Q}QV^{^\prime}P
\end{align}
From (\ref{hp01}) and (\ref{p}), the first term in (\ref{heff2}) can be shown to be
\begin{align}
\label{heft1}
PH^{^\prime}_0P=(E_1-\hbar\omega_1)\mid e_1\rangle\langle e_1\mid+(E_2-\hbar\omega_2)\mid e_2\rangle\langle e_2\mid+E_3\mid c\rangle\langle c\mid
\end{align}
Using (\ref{vp1}) and (\ref{p}), The second term in (\ref{heff2}) comes out to be
\begin{align}
\label{heft2}
PV^{^\prime}P=\hbar\Omega^{^\prime}_{13}\mid e_1\rangle\langle c\mid+\hbar\Omega^{^\prime}_{23}\mid e_2\rangle\langle c\mid
\end{align}
The deduction of the last term in (\ref{heff2}) is cumbersome and make use of Eq. (\ref{hp01}-\ref{q}) and gives
\begin{align}
\label{heft3}
&PV^{^\prime}Q\frac{1}{z-QH^{^\prime}Q}QV^{^\prime}P\nonumber\\[16pt]=4\Bigg[\int\limits_0^\infty dk&\frac{\left|V^{^\prime}_{1k}\right|^2}{z-k}\mid e_1\rangle\langle e_1\mid+\int\limits_0^\infty dk^{^\prime}\frac{\left|V^{^\prime}_{2k^{^\prime}}\right|^2}{z-k^{^\prime}}\mid e_2\rangle\langle e_2\mid+\int\limits_0^\infty dk\frac{V^{^\prime *}_{1k}V^{^\prime}_{2k^{^\prime}}}{z-k}\mid e_1\rangle\langle e_2\mid\nonumber\\&+\int\limits_0^\infty dk^{^\prime}\frac{V^{^\prime}_{1k}V^{^{\prime *}}_{2k^{\prime}}}{z-k^{^\prime}}\mid e_2\rangle\langle e_1\mid\Bigg]\nonumber\\
+\int\limits_0^\infty dE\Bigg[&\frac{\left|\Lambda^{^\prime}_{1E}\right|^2}{z-E}\mid e_1\rangle\langle e_1\mid+\frac{\left|\Lambda^{^\prime}_{2E}\right|^2}{z-E}\mid e_2\rangle\langle e_2\mid+\frac{\left|V^{^\prime}_{3E}\right|^2}{z-E}\mid c\rangle\langle c\mid+\frac{\Lambda^{^\prime}_{1E}\Lambda^{^{\prime *}}_{{2E}}}{z-E}\mid e_1\rangle\langle e_2\mid\nonumber\\
&+\frac{\Lambda^{^\prime}_{2E}\Lambda^{^{\prime *}}_{{1E}}}{z-E}\mid e_2\rangle\langle e_1\mid+\frac{\Lambda^{^\prime}_{2E}V^{^{\prime *}}_{3E}}{z-E}\mid e_2\rangle\langle c\mid+\frac{V^{^\prime}_{3E}\Lambda^{^{\prime *}}_{{2E}}}{z-E}\mid c\rangle\langle e_2\mid\nonumber\\
&
+\frac{\Lambda^{^\prime}_{1E}V^{^{\prime *}}_{3E}}{z-E}\mid e_1\rangle\langle c\mid+\frac{V^{^\prime}_{3E}\Lambda^{^{\prime *}}_{{{1E}}}}{z-E}\mid c\rangle\langle e_1\mid\Bigg]
\end{align}
The Eq. (\ref{heft3}) consists of two parts. The terms which contain the $k$-integrals represent the effect of vacuum interaction. We consider it as the first part. The last nine terms constitute the second part which represents the effect of the applied lasers and the magnetic field. We first do the $k$-integrals having simple pole structure. For that we consider $k$ to be slightly complex $k\rightarrow k+i\epsilon$ and use the prescription $\lim\limits_{\epsilon\to 0}\frac{1}{k-z-i\epsilon}={\cal P}\frac{1}{k-z}+i\pi\delta(k-z)$, where $\cal P$ denotes Cauchy principal value. The vacuum coupling parameters $V^{^\prime}_{1k}$ $\left(V^{^\prime}_{2k^{^\prime}}\right)$ can be assumed to have a significant value very close to $k=k_{1f}$ $\left(k_{2f}\right)$, the energy difference between the electronic states $\mid f\rangle$ and $\mid e_1\rangle$ $\left(\mid e_2\rangle\right)$. So, we can replace $V^{^\prime}_{1k}$ $\left(V^{^\prime}_{2k^{^\prime}}\right)$ by $V^{^\prime}_{1f}$ $\left(V^{^\prime}_{2f}\right)$ and take them outside the integrals. On doing so, the principal part 
becomes zero. Thus, in this way we neglect all the vacuum induced shifts and we are then left only with the vacuum induced widths from the imaginary part of the $k$-integrals. Since the vacuum interaction is very weak, this can be considered a good approximation.
The width coming from the imaginary part of the diagonal terms may be identified as the spontaneous decay rates.
\begin{align}
\gamma_{1}=\frac{4\pi}{\hbar}\left|V^{^\prime}_{1f}\right|^2\hspace{1cm}\gamma_{2}=\frac{4\pi}{\hbar}\left|V^{^\prime}_{2f}\right|^2
\end{align}
$\gamma_{_{VIC}}$ represent vacuum induced coherence (VIC) coming from the off-diagonal terms given by
\begin{align}
\label{eta12}
\gamma_{_{VIC}}=\frac{2\pi}{\hbar}V^{^{\prime *}}_{1f}V^{^\prime}_{2f}
\end{align}
We have assumed that both the lasers have the same phase and so that $\gamma_{_VIC}$ is independent of laser phase. It can be shown that this term only exists if the two continuum $\mid f,k\rangle$ and $\mid f,k^{^\prime}\rangle$ becomes non-orthogonal for some value of $k$.

The second part of Eq. (\ref{heft3}) contains only $E$-integration and gives rise to laser and magnetic field induced shifts and widths as well as coherence between them. Similar to the prescription done in doing $k$-integration we here consider $E$ to be slightly complex $E\rightarrow E+i\epsilon$ and use $\lim\limits_{\epsilon\to 0}\frac{1}{E-z-i\epsilon}={\cal P}\frac{1}{E-z}+i\pi\delta(E-z)$. If we consider the presence of a Feshbach resonance at $E=E_3$, then the bound-continuum coupling terms $\Lambda^{^\prime}_{1E}$, $\Lambda^{^\prime}_{2E}$ as well as $V^{^\prime}_{3E}$ assume sharp variation in the neighbourhood of $E_3$. Thus, significant contribution to the integrals come from that neighbourhood and we can replace $z$ by $E_3$. We sort out some important terms as follows. $E^{^{\rm{sh}}}_{1}$, $E^{^{\rm{sh}}}_{2}$ and $E^{^{\rm{sh}}}_F$ are shifts of the eigenvalues of $H^{^\prime}_0$ and are given by
\begin{align}
E^{^{\rm{sh}}}_{1}=&{\cal P}\int\limits_0^\infty dE\frac{\left|\Lambda^{^\prime}_{1E}\right|^2 }{E_3- E}
\end{align}
\begin{align}
E^{^{\rm{sh}}}_{2}=&{\cal P}\int\limits_0^\infty dE\frac{\left|\Lambda^{^\prime}_{2E}\right|^2}{E_3-E}
\end{align}
\begin{align}
E^{^{\rm{sh}}}_F={\cal P}\int\limits_0^\infty dE\frac{\left|V^{^\prime}_{3E}\right|^2}{E_3- E}
\end{align}
$\beta_{1}$, $\beta_{2}$ and $\alpha$ are complex quantities which appear in the off-diagonal part and are given by
\begin{align}
\beta_n={\cal P}\int\limits_0^\infty dE \frac{\Lambda^{^\prime}_{nE}V^{^{\prime *}}_{3E}}{E_3-E}
\end{align}
\begin{align}
\alpha&={\cal P}\int\limits_0^\infty dE \frac{\Lambda^{^\prime}_{1E}\Lambda^{^{\prime *}}_{2E}}{E_3-E}
\end{align}
The laser induced widths $\Gamma_{1}$, $\Gamma_{2}$ and the width $\Gamma_{F}$ of the FBR are given by
\begin{align}
\Gamma_{1}=\frac{2\pi}{\hbar}\left|\Lambda^{^\prime}_{1F}\right|^2 \hspace{1cm}\Gamma_{2}=&\frac{2\pi}{\hbar}\left|\Lambda^{^\prime}_{2F}\right|^2\hspace{1cm} \Gamma_{F}=\frac{2\pi}{\hbar}\left|V^{^\prime}_F\right|^2
\end{align}
where $\Lambda^{^\prime}_{1F}$, $\Lambda^{^\prime}_{2F}$ and $V^{^\prime}_F$ are the values of $\Lambda^{^\prime}_{1E}$, $\Lambda^{^\prime}_{2E}$ and $V^{^\prime}_{3E}$ respectively, calculated at $E=E_3$.
$\gamma_{_{LIC}}$ represents the coherence between the two lasers which is given by
\begin{align}
\label{G12}
\gamma_{_{LIC}}=\frac{2\pi}{\hbar}\Lambda^{^\prime}_{1F}\Lambda^{^{\prime *}}_{2F}
\end{align}
$\Gamma_{1f}$ and $\Gamma_{2f}$ represents cross coupling between the lasers and the magnetic field 
\begin{align}
\label{Gnf}
\Gamma_{nF}=\frac{2\pi}{\hbar}\Lambda^{^\prime}_{nF}V^{^{\prime *}}_F
\end{align}
For the sake of calculational simplicity, we assume $\Gamma_{nF}$, $\gamma_{_{LIC}}$, $\gamma_{_{VIC}}$, $\alpha$, $\beta$ and $\Omega_{n3}$ are real quantities.

\end{document}